\documentclass{article}

\usepackage{microtype}
\usepackage{graphicx}
\usepackage{subfigure}
\usepackage{booktabs} 

\usepackage{hyperref}
\usepackage{aas_macros}

\usepackage[accepted]{icml2022}

\usepackage{amsmath}
\usepackage{amssymb}
\usepackage{mathtools}
\usepackage{amsthm}

\icmltitlerunning{Galaxy Merger Reconstruction with Equivariant Graph Normalizing Flows}

\begin{document}

\twocolumn[
\icmltitle{Galaxy Merger Reconstruction with Equivariant Graph Normalizing Flows}



\icmlsetsymbol{equal}{*}

\begin{icmlauthorlist}
\icmlauthor{Kwok Sun Tang}{equal,to}
\icmlauthor{Yuan-Sen Ting}{equal,rsaa,soco}
\end{icmlauthorlist}

\icmlaffiliation{to}{Department of Astronomy, University of Illinois Urbana-Champaign, USA}
\icmlaffiliation{rsaa}{Research School of Astronomy \& Astrophysics, Australian National University, Cotter Rd., Weston, ACT 2611, Australia}
\icmlaffiliation{soco}{School of Computing, Australian National University, Acton, ACT 2601, Australia}

\icmlcorrespondingauthor{Kwok Sun Tang}{kwoksun2@illinois.edu}
\icmlcorrespondingauthor{Yuan-Sen Ting}{yuan-sen.ting@anu.edu.au}

\icmlkeywords{Machine Learning, ICML, astronomy, astrophysics}

\vskip 0.3in
]



\printAffiliationsAndNotice{\icmlEqualContribution}

\begin{abstract}
A key yet unresolved question in modern-day astronomy is how galaxies formed and evolved under the paradigm of the $\Lambda$CDM model. A critical limiting factor lies in the lack of robust tools to describe the merger history through a statistical model. In this work, we employ a generative graph network, E(n) Equivariant Graph Normalizing Flows Model. We demonstrate that, by treating the progenitors as a graph, our model robustly recovers their distributions, including their masses, merging redshifts and pairwise distances at redshift $z=2$ conditioned on their $z = 0$ properties. The generative nature of the model enables other downstream tasks, including likelihood-free inference, detecting anomalies and identifying subtle correlations of progenitor features.
\end{abstract}

\section{Introduction}
\label{intro}

The standard $\Lambda $CDM cosmological model has predicted the hierarchical structure formation; smaller galaxies merge throughout cosmic history to form the present-day galaxies. This is further supported by hydrodynamic simulations, which demonstrated that the merging history critically determines the emergence of the galaxies and their properties \cite{Wechsler_2018}. While we have a broad-brush understanding of how galaxies evolve, a quantitative understanding remains elusive. In recent years, modeling how galaxy progenitors impact its formation has received much-renewed interest. On the one hand, data from \textit{GAIA} satellite has enabled astronomers to fathom the accretion events of our Milky Way galaxy \citep{Helmi2020}. On the other hand, the newly launched James Webbs Space Telescope is poised to revolutionalize our understanding of galaxy formation during the cosmic dawn \cite{2018MNRAS.474.2352C,2019MNRAS.483.2983Y, Behroozi_2020}.

The quest to understand the connection between the near-field cosmology and the high-redshift cosmic past has led to many semi-analytical or empirical models to infer the galaxy properties from their merger history. For example, those models have been employed in-painting dark matter halo only simulations with baryonic properties in large-scale cosmological simulations \cite{kamdar2016a, kamdar2016b, jokim2019, Lovell2021}. However, most of these classical approaches rely on studying the global statistical connection between dark matter haloes and galaxies, often reducing the study to focusing on the connection of individual haloes and galaxies or summarizing their formation environment with crude summary statistics based on human heuristics.

\begin{figure*}\label{fig:schematic}
\begin{center}
    \includegraphics[width=0.85\linewidth]{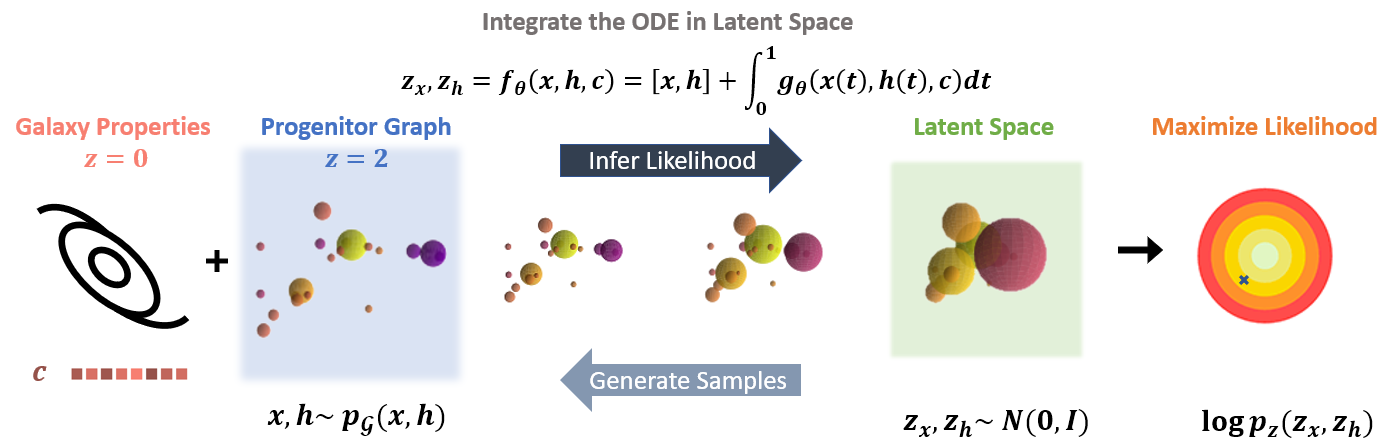}
    \caption{Generative Graph Normalizing Flows as a way to model merger histories of galaxies. We adopt E(n) Equivariant Normalizing Flows to model the conditional distribution of progenitor graphs at redshift $z=2$ given the observable quantities at $z=0$.}
\end{center}
\vskip -0.2in
\end{figure*}

The advancement in graph neural networks (GNN) \cite{Bruna2013,Defferrard2016,KipfWelling2016} has opened up many new possibilities for studying the evolutionary history of the galaxies. That is because a graph is a natural descriptor of the systems at hand -- any progenitor system at a high redshift can be regarded as a graph, with individual progenitors as nodes on the graph (see Fig.~\ref{fig:schematic}). Noting this connection, various works have started to apply GNN to astronomy \cite{cranmer2020,gnn_halomass,gnn_alignment}. However, almost all of these GNN methods employed thus far are discriminative, i.e., it infers the labels from the graph. While it has much inroads, the discriminative nature of these works somewhat limits the true power of describing galaxy evolutionary history through graphs.

To bridge this gap, we propose using generative graph networks to study the merger history of galaxies. Unlike the classical approach, graphs allow us to incorporate the progenitors' spatial information and extend our merger history reconstruction to an arbitrary number of nodes. Furthermore, our approach allows us to perform multiple downstream tasks, such as likelihood-free inference and out-of-distribution detection, going beyond discriminative GNNs.

\section{Cosmological Simulations}
\label{data}

This study will explore the reconstruction of the progenitor graph in the early Universe (at redshift, $z=2$) through the present day ($z=0$) observables. We will train our model on the cosmological simulations from \texttt{TNG300} \cite{Nelson2019}. We extract halo merger trees and limit our current study to subhalos at $z=0$ with a mass between $0.5 - 2.5  \times 10^{12} {\rm M}_{\odot}$. This leaves us with 76105 halos. We will treat the halo properties at $z = 0$ as ``conditional vectors". In this current study, these are limited to eight variables: the black hole mass, black hole accretion rate, gas metallicity, star metallicity, star formation rate, maximum velocity of the spherically averaged rotation curve, and velocity dispersion and their respective mass. As for the progenitor graph, we consider the masses, positions and merging redshifts of the respective progenitors at $z = 2$.

\section{Generative Graph Normalizing Flows}
\label{model}

Our model is based on the E(n) Normalizing Flows \cite{en_flows} (or E(n) Flow in short), a recently proposed machine learning model that aims at simultaneously generating molecule features and their 3D positions as a graph. We refer readers for the detailed implementation in \citet{en_flows}. Here we only outline the major components, and the schematic of our method is shown in Fig.~\ref{fig:schematic}.

In our problem setup, each progenitor graph with $M$ halos is specified with the two node features (mass and merging redshift) $\mathbf{h} \in R^{M \times 2}$ and their position coordinates $\mathbf{x} \in R^{M \times 3}$. The nodes are ordered by mass and are labeled in the subscript to reflect its ordering, i.e. $M_n$, $z_n$  corresponds the mass and merging redshift of the $n^{th}$ most massive progenitor (``0" denotes the main parent progenitor). $d_{ij}$ corresponds to the relative distance between $i^{th}$ and $j^{th}$ progenitor. E(n) Flow aims to transform the distribution of all progenitor graphs $p_{\mathcal{G}} (\mathbf{x}, \mathbf{h})$ in to a tractable base distribution
$p_Z (\mathbf{z_x}, \mathbf{z_h})$. The transformation is done through an invertible neural network transformation $f_\theta$, which we optimize through maximum likelihood. More specifically, given the ensemble of progenitor graphs in \texttt{TNG300}, the likelihood of each graph $\mathbf{x_i}, \mathbf{h_i}$ can be evaluated using the change of variables formula, where $
    p_\mathcal{G} (\mathbf{x_i}, \mathbf{h_i}) = p_Z ( f_\theta(\mathbf{x_i}, \mathbf{h_i}) ) ~ \rm{det} \left| \frac{\partial f_\theta}{\partial G} \right|.
$

\begin{figure}[t]
\begin{center}
    \includegraphics[width=\linewidth]{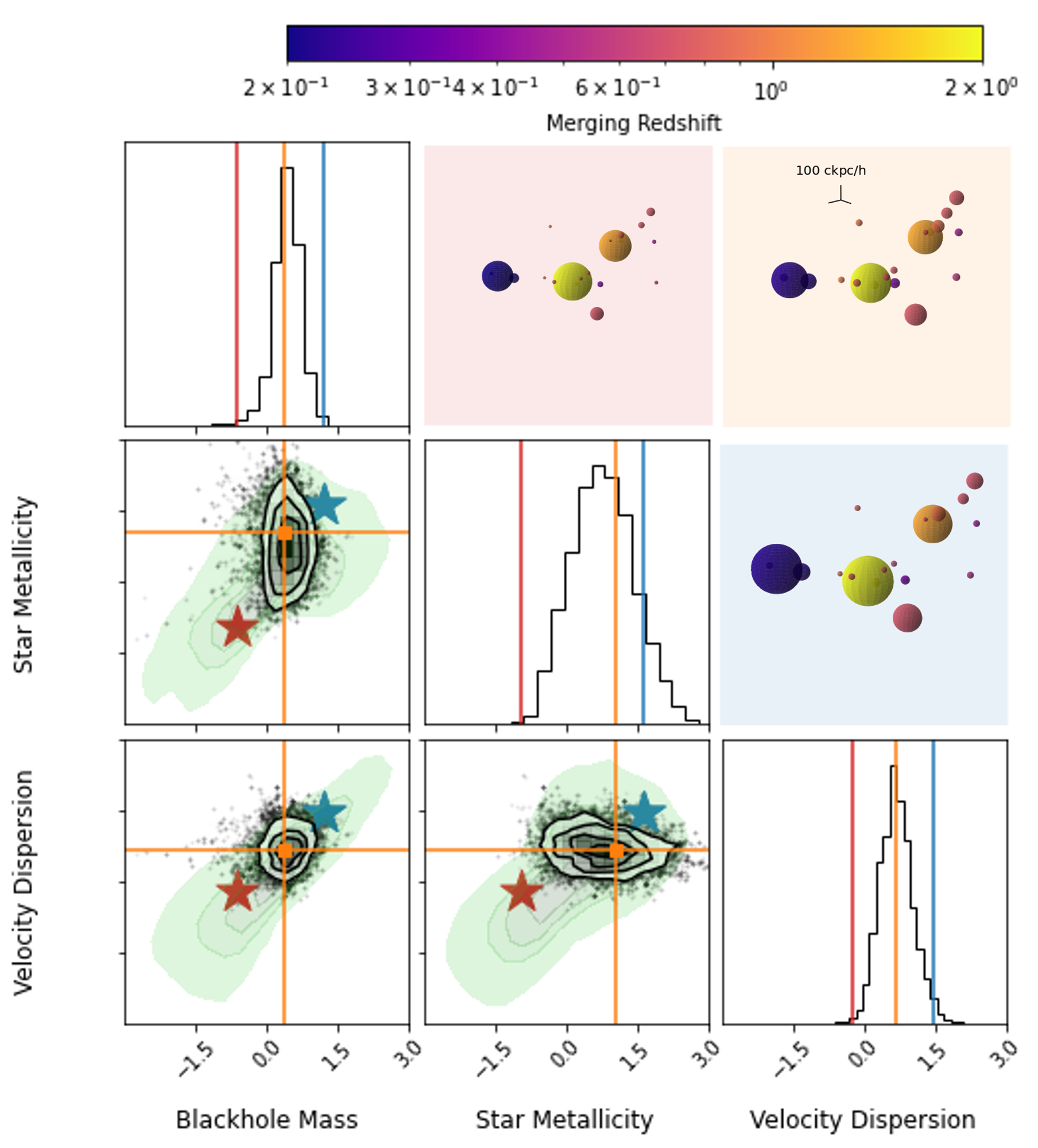}
    \vspace{-0.5cm}
      \caption{Likelihood-free inference with E(n) Flow. The lower triangular panels show the posterior probability distribution $p(c|\tilde{G})$, with the target progenitor graph $\tilde{G}$ shown in top right corner. The orange crosshair shows the ground truth parameters. While our sampling works in all eight conditional observables, we only show a 3D marginal posterior for illustrative purposes. The underlying distribution of the conditional observables $p(c)$ is shown in solid green contours. The red and blue panels show progenitor graphs sampled from $p_\theta(G|c)$. The size of each node denotes the progenitor mass and the color of the merging redshift. The red and blue stars in the lower triangular plots are the conditional vectors sampled to generate the the progenitor graphs shown in the red and blue upper panels.
    }
    \label{fig:mcmc}
\end{center}
\vskip -0.2in

\end{figure}

In the case of E(n) flow, the invertible transformation is assumed to be an FFJORD \cite{ffjord}. Briefly, FFJORD is a continuous-time normalizing flow model that maps the latent to data distribution with a neural ODE $g_\theta$ where $\mathbf{z} = \mathbf{x(0)} + \int_0^1 g_\theta (\mathbf{x(t)}) dt$. The model allows for a straightforward invertible transform by integrating backward in time, i.e. $\mathbf{x} = \mathbf{z(0)} + \int_1^0 g_\theta (\mathbf{x(t)}) dt$, and hence allowing the generation of graphs from $p_{\mathcal{G}} (\mathbf{x}, \mathbf{h})$.

The challenge of training any graph-based neural network always has an enormous implicit degree of freedom which exacerbate the curse of dimension. E(n) Flow tackles this problem by imposing, as the name suggests, an E(n) group equivalence. E(n) is the $n$-dimensional Euclidean group where transformations of this group preserve Euclidean distances, including translations, rotations, and reflections. The E(n) is particularly relevant for this study due to the same inherent symmetries progenitor graphs possess. In E(n) Flow, to enforce equivariance, $g_\theta$ is chosen as the Equivariant Graph Neural Network (EGNN) \citep{egnn}. Briefly, EGNN consists of layers of Equivariant Graph Convolutional Layer (EGCL). The edge function $\phi_e$ in EGCL takes the relative squared distance between the nodes as additional input when edge embedding between the nodes is evaluated. The coordinates of each node are updated by a weighted sum of all relative displacement between the nodes, leading to the E(n) equivariance.

Importantly, since $g_\theta$ operates on a graph with an arbitrary number of nodes, the model allows input graphs for varying $M$, where $M$ is the number of nodes. Through message passing within $g_\theta$, the progenitor graph with different $M$ will be mapped to the corresponding $p_Z (\mathbf{z_x}, \mathbf{z_h})$ and evaluate the likelihood. And vice versa, during the generation process, we can draw $(\mathbf{z_x}, \mathbf{z_h})$ from any fixed $M$ and run the inverse transform accordingly.

Note that the primary goal of this study is to study progenitor graphs at $z=2$, given the conditional observable at $z=0$. Therefore, instead of training the native formulation proposed in \citet{en_flows}, we modify the E(n) flow to capture also the conditional distribution. This is done by incorporating the conditional vector as additional input for all the edge aggregation functions in EGCL.

Finally, our codes are made publicly available on Github\footnote{\url{https://github.com/hisunnytang/HaloEGNN}}. In our case, training on a single NVIDIA-A100 converges after 192 hours. AdamW optimizer is used with a \texttt{ReduceOnPleatau} scheduler that reduces the learning rate by a factor of $10$ when the metric has stopped improving for a patience of three epochs. Training is stopped when the validation loss stops improving for five epochs.

\section{Results}
\label{results}

The ability to summarize progenitor graphs via a robust statistical framework enables many new opportunities. Below we will highlight three key applications of E(n) Flow.
\begin{figure}[t]
\begin{center}
    \includegraphics[width=0.9\linewidth]{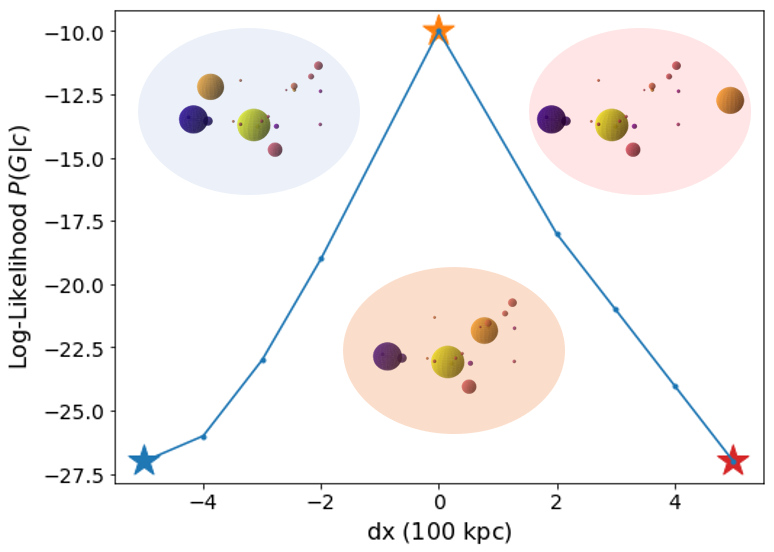}
    \caption{Out-of-distribution detection. As we perturb the location of the second-most massive progenitor (the orange node), the likelihood decreases, showing that such configurations are deemed less likely by E(n) Flow than the unperturbed ground truth.
    }
    \label{fig:outlier}
\end{center}
\vskip -0.2in
\end{figure}

{\bf Likelihood-Free Inference:} A myriad of decadal astronomical flagships, including the Rubin Observatory, James Webbs Space Telescope, and Roman Space Satellite, will come to fruition this decade, drastically revolutionizing the study of galaxy evolution. A critical frontier is to connect the observations at high redshifts and infer their counterparts in the Local Universe. Our model provides a unique opportunity to tackle this question.

In particular, our trained E(n) Flow $p_\theta (G|c)$ can serve as the likelihood function which allows us to query the posterior estimate $p(c|G) \propto p_\theta(G| c) p (c)$. As a proof of concept, Fig.~\ref{fig:mcmc} shows a case study where a specific target progenitor graph $\tilde{G}$ is given (shown in the orange panel). We sample the posterior $p(c|\tilde{G})$ with MCMC, and the 1, 1.5, 2$\sigma$ of the posterior is shown in contours in the lower triangular plots. Fig.~\ref{fig:mcmc} demonstrates that with E(n) Flows, we can robustly infer the projected low-redshift properties directly from the high-redshift progenitor graph. Also shown in green is the support of the conditional observable distribution $p(c)$ from the entire data distribution. Comparing the contours with the shaded background further demonstrates the constraining power of observing the progenitor graph at high-redshift.

To gain further intuition on how the inference was determined, we generated two proposed graphs (in the red and blue panels) with the same latent vectors $z$ that corresponds to $\tilde{G}$. The progenitor graph corresponding to the blue conditional vector shows more massive progenitor distributions, and the progenitors are more spread out in space. The progenitor graph in red indicates a slightly less massive $M_1$ instead of almost equal mass $M_0$ and $M_1$ in our target graph. These generations illustrate why the conditional vectors corresponding to these configurations are considered possible ($3\sigma$) but far less probable than the ground truth.

\begin{figure}[t]
\begin{center}
    \includegraphics[width=1\linewidth]{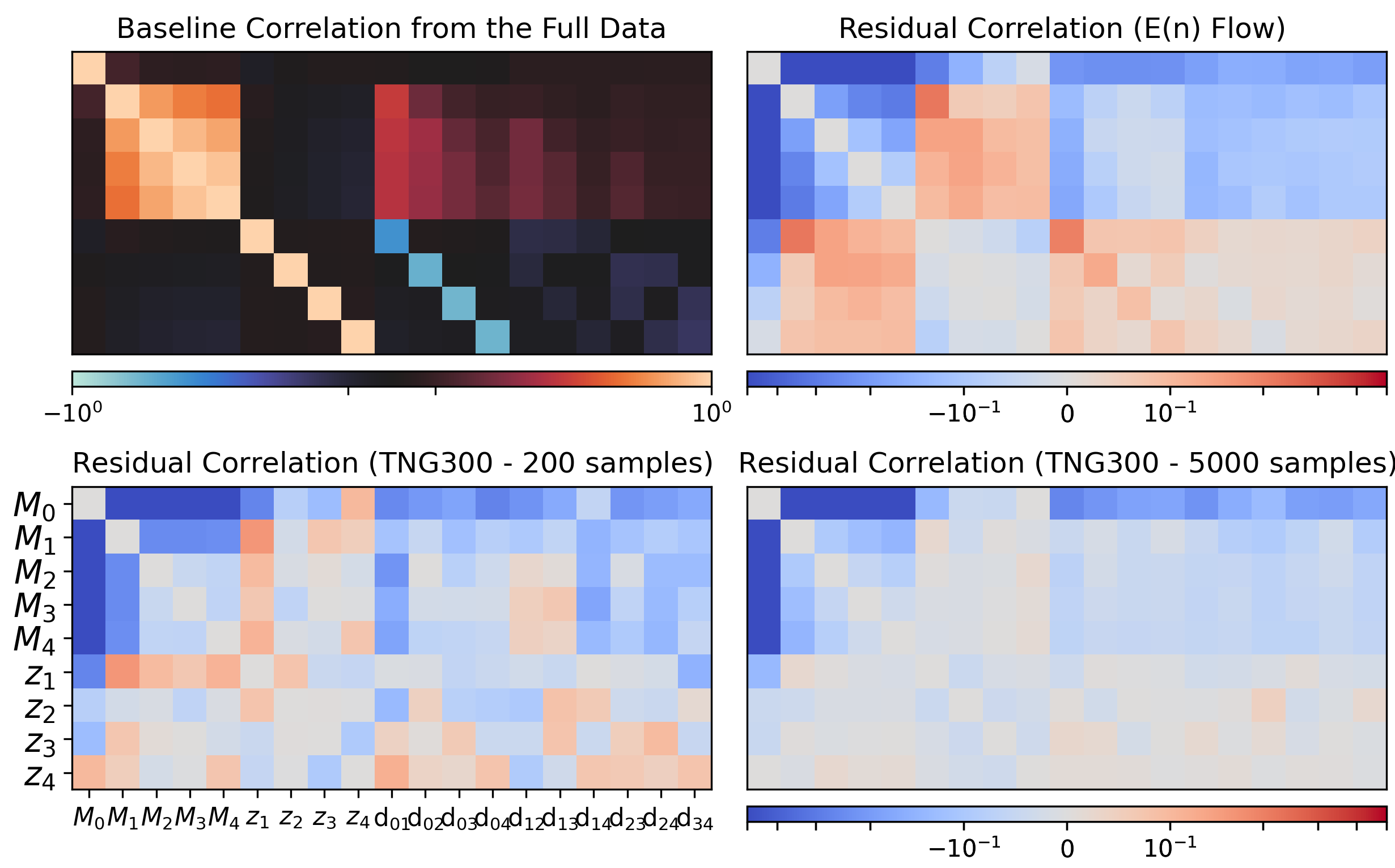}
    \caption{E(n) Flow can identify subtle correlation between the progenitors features. Upper left panel shows the correlation from the full dataset. Upper right panel shows the residual correlation at a fixed conditional vector $\tilde{c}$ from E(n) Flow. Lower left and right panels shows respectively the residual correlation generated from 200 and 5000 progenitor graphs with the closest $c\simeq \tilde{c}$ in \texttt{TNG300}.
    }
    \label{fig:corr}
\end{center}
\vskip -0.2in
\end{figure}

{\bf Out-of-Distribution Detection:} The vast sample from large surveys (e.g., Rubin, Roman) is bound to find many unexpected outliers. Unlike discriminated GNNs, the generative approach proposed here comes with a critical advantage in looking for galaxies with uncanny configurations.

To demonstrate how we can detect outliers through graph configuration, Fig.~\ref{fig:outlier} examines how the likelihood $p_\theta (G | c)$ varies as we change the property of a progenitor galaxy in the graph. In this specific exploration, we move the second most massive progenitor (shown in orange) along the x-direction. The unperturbed graph achieves the highest log-likelihood, meaning that this overall graph configuration is deemed possible by E(n) Flow. However, as one of the progenitors is translated in its spatial location, the log-likelihood decreases. This demonstrates that E(n) Flow is sensitive to configurations unseen in \texttt{TNG300}.

{\bf Correlation Identification:} Summarizing all progenitor graphs through a robust statistical model $p_\theta (G | c)$ further enables us to unearth subtle correlations in galaxy evolution that would otherwise be inaccessible to us by simply "slicing" the cosmological simulations. Even with the entire simulated dataset, there is no more than one progenitor graph $G$ realization with any vector $\tilde{c}$. Consequently, to probe the correlations between the progenitor features at a fixed conditional $\tilde{c}$, one could usually only resort to sampling the progenitors' graphs with conditional vectors in the neighborhood of $\tilde{c}$ that one is interested in. E(n) Flow offers a unique way to recover the correlations between the progenitor features by repeatedly sampling from the flow model while keeping the conditional vector $\tilde{c}$ fixed.

We first illustrate the progenitor feature correlations of the entire dataset $p_\mathcal{G}(G)$ (i.e., integrating over all conditional vectors) in the top panel of Fig.~\ref{fig:corr}. There are notable correlations between various progenitor features even without conditioning on the conditional vectors. Unfortunately, these baseline correlations can often render other subtle correlations invisible when conditioning on the conditional vectors. And this is well illustrated in the other panels in Fig.~\ref{fig:corr}.

The bottom panel show the residual correlation plot generated from 200 and 5000 progenitor graphs with the closest conditional vector $\tilde{c}$ with a black hole mass of $\rm 8.4 \times 10^{7} M_{\odot}/h$, blackhole accretion rate of $\rm 1.7 \times 10^{7} M_{\odot}/ Gyr$, gas metallicity of $0.015$, star metallicity of $0.021$, mass of $\rm 5.5 \times 10^{11} M_{\odot}$, star formation rate of $\rm 2.08 ~ M_{\odot}/ yr$, maximum velocity of $\rm 187 ~ km/s$ and velocity dispersion of $\rm 108 ~ km/s$. The residual correlation is defined as the observed correlation subtracting the baseline correlations. On the one hand, even with a cutting-edge cosmological simulation like \texttt{TNG300}, a smaller sample shows a noticeable yet noisy residual correlation. On the other hand, taking significantly more progenitor graphs in the neighbourhood of $\tilde{c}$ can indeed bring down the sampling noise. However, this is at the expense of the correlation, as this large ensemble of conditional variables deviates more from $\tilde{c}$ and thus wipes out the residual correlations.

The E(n) Flow reveals a much clearer residual correlation. The residual correlation was evaluated through repeated sampling from $p_\theta (G | \tilde{c})$. Consistent with what has already been alluded in Fig.~\ref{fig:mcmc} (hollow contours vs. green contours), the features of $G$ in $p_\theta(G|\tilde{c})$ can exhibit different correlations (and sometimes even different sign) from $p_\mathcal{G}(G)$.

\section{Concluding Remarks}

Untangling galaxy evolution is a messy business, yet it underlines the cosmic evolution that subsequently led to our own existence. In recent years, cutting-edge cosmological simulations have allowed us to pierce into the complex physical processes that engender galaxy evolution. The simulations further complement the many exciting data sets from the astronomical flagships that have come to fruition this decade (Euclid, Rubin, Roman). However, what has been critical is missing a statistically robust way to describe galaxies as an ensemble and encode their connections.

This study demonstrated that generative graph normalizing flows might hold the key to resolving this perennial conundrum. While we apply to a specific case study -- connecting high-redshift progenitor graphs with the local observables, our method can be extended naturally to other domains. For example, the same idea can also be used to model the lens plane of strong lensing sources by conditioning on the unresolved properties of the lens galaxies. It can model the locally observed graph, potentially shedding light on the still debated plane of satellites in the Milky Way and M31 \cite{2018MPLA...3330004P,sawala2022milky}.

In a nutshell, a generative graph normalizing flows allows us to put galaxy evolution on a more robust statistical footing. And that might finally ``connect the dots" and shed light on the ultimate question of where we come from.

\bibliography{main}
\bibliographystyle{icml2022}





\end{document}